\documentstyle[aps,preprint,epsfig]{revtex} 

\begin{document}

\draft

\preprint{nucl-th/0108049}


\title{Dynamical Study of the $d(e,e' \pi^+)$ reaction}


\author{ K. Hafidi and T.-S. H. Lee }


\address{
Physics Division, Argonne National Laboratory, Argonne,
Illinois 60439}


\begin{abstract}
The $d(e,e'\pi^+)$ reaction in the parallel kinematics
has been investigated using
a dynamical model of pion electroproduction on the nucleon.
A unitary $\pi NN$ model has been used in order to examine the effects
due to the final two-nucleon interactions, pion rescattering from the
second nucleon, and the intermediate $NN$ and $N\Delta$ interactions.
It has been found that these $\pi NN$ mechanisms are
small, but they
can have significant contributions
to the $d(e,e'\pi^+)$ cross sections through their interference
with the dominant impulse term. For the longitudinal cross sections,
the effects due to the interference between the
pion pole term and
other production mechanisms are also found to be
very large. Our findings clearly indicate that these interference 
effects must be accounted for in any attempt to determine 
from the $d(e,e'\pi^+)$ data
whether the pion form factor and/or $\pi NN$ vertex of the pion pole
term are modified in the nuclear medium.

\end{abstract}

\maketitle

\section{Introduction}

It has been suggested that the pion form factor and/or the
$\pi NN$ vertex could be modified by the nuclear medium. One possible
way to investigate this interesting question is to study the 
electroproduction of charged pions on the nuclei in the kinematic region
where the outgoing pions are in the direction of the exchanged
virtual photon. For this so-called parallel kinematics, 
the pion pole term (i.e the pion-exchange production
amplitude of Fig.\ref{fig2}a) is expected to be dominant and hence the measured cross 
sections could be used to explore the medium effects on the pion form 
factor and/or the $\pi NN$ vertex. 
The above consideration has motivated an experiment
 at Saclay in 1990 \cite{saclay}
and another \cite{anl} at
Jefferson Laboratory (JLab). The objective
of this work is to address some theoretical questions concerning
the interpretation of the data from these experimental efforts.
Here, we consider the simplest $d(e,e'\pi^+)$ reaction.

We use the dynamical model developed by Sato 
and Lee \cite{sl} (called the SL model).
The aim of the SL model is to interpret the data of pion photoproduction
and electroproduction  in terms of quark sub-structure of hadrons.
The outcome of this effort has two aspects : (i) 
to establish the interpretation 
of the $\gamma N\rightarrow \Delta$ excitation in terms of constituent 
quark models. (ii) 
to have a dynamical model which
gives a fairly accurate description of all of the data of pion
photoproduction and electroproduction, which can be used to
perform various nuclear calculations. We make use of the second aspect 
of the SL model in this investigation.

The SL model is illustrated in Fig.\ref{fig1}. It consists
of a production term, illustrated in Fig.2,
and a term involving $\pi N$ scattering.
The focus of the experiments with parallel kinematics is
the pion pole term, Fig.\ref{fig2}a, which depends on the pion form factor and
the $\pi NN$ vertex. Clearly, the effects due to the other terms 
in Figs.\ref{fig1}-\ref{fig2} must be carefully investigated before the medium effects
on the pion pole term can be determined. This will be the ultimate goal of 
this work.

The second problem we want to address is the extent to which the
effect due to the pion pole term will be masked by the hadronic 
final state interactions. For the $d(e,e'\pi^+)$ reaction, the
final $\pi NN$ interactions can in principle be calculated by using the
unitary formulation developed in Refs.\cite{garcilazo,lm}. This
is a rather complex numerical task and will not be pursued in this work.
Instead we consider only 
the leading terms of a multiple scattering
expansion of the scattering amplitude defined by the $\pi NN$ model
of Ref.\cite{lm}.   

We thus calculate  
the $d(e,e'\pi^+)$ cross sections from the four leading
mechanisms illustrated in Fig.\ref{fig3}.
The impulse term($Imp$), illustrated in
Fig.\ref{fig3}a, is due to the production on a nucleon in the deuteron.
The other three terms are due to final two-nucleon interactions,
pion rescattering from the second nucleon ($\pi NN$), and intermediate $NN$ and
$N\Delta$ interactions. The amplitudes of these reaction mechanisms can
be evaluated straightforwardly from using the SL model and the $\pi NN$
model developed in Refs.\cite{lm}.
Thus our calculations of the $d(e,e'\pi^+)$ cross sections
will be free of adjustable parameters.

In section II, we present the formulation for calculating
the cross sections of $d(e,e'\pi)$ reaction. The results are
presented and discussed in section III.

\section{Formulation}

In the rest frame of the initial 
deuteron, the differential cross section of the
$d(e,e'\pi)$ reaction can the be calculated from
\begin{eqnarray}
\frac{d^{6}\sigma }{d\Omega _{e'}dE_{e'}d\Omega _{\pi }dk} 
~&& = \sigma _{M}\left[ W_{2}-2W_{1}\tan ^{2}(\frac{\theta _{e}}{2})\right]\nonumber
\\
&& = \Gamma _{v} \left[ \sigma _{T}+\varepsilon \sigma _{L}\right] ,
\end{eqnarray}
where 
 $\theta_e$ is the angle between the outgoing and incoming electrons,
$E_{e'}$ is the outgoing electron energy, and $\vec{k}$ is the outgoing pion 
momentum.
The Mott cross section is defined by
\begin{equation}
\sigma _{M}=\frac{\alpha ^{2}\cos ^{2}
(\frac{\theta _{e}}{2})}{4E_{e}\sin ^{4}(\frac{\theta _{e}}{4})}.
\end{equation}
where $\alpha=\displaystyle{\frac{e^2}{4\pi}}=\displaystyle{\frac{1}{137}}$. If the photon 
four-momentum is denoted as $q=(\omega,\vec{q})$, the other kinematic 
factors in Eq.(1) are 
\begin{eqnarray}
\varepsilon &=&\frac{1}
{1+2\frac{\left| \vec{q}\right| ^{2}}{Q^{2}}
 \tan ^{2}(\frac{\theta _{e}}{2})}, \\
\Gamma _{v}&=&\frac{\alpha }{2\pi ^{2}}\frac{E_{e'}}{E_{e}}
\frac{K}{Q^{2}}\frac{1}{1-\varepsilon }. 
\end{eqnarray}
Here we have defined $Q^2=-q^2= \mid \vec{q}\mid ^2 - \omega^2$ 
and  $ K=\displaystyle{\frac{W^{2}-m_d^{2}}{2m_d}}$ with $m_d$ being 
the deuteron mass and $W=[(\omega+m_d)^2-\vec{q}^2]^{1/2}$ being 
the invariant mass of the $\gamma d$ system. 
From the above definitions, one can
show that the longitudinal cross section $\sigma_L$ and 
transverse cross section $\sigma_T$ in Eq.(1) are related to the
structure functions $W_1$ and $W_2$ by
\begin{eqnarray}
\sigma _{T}&=&\frac{4\pi ^{2}\alpha }{K}\left[ -W_{1}\right], \\
\sigma _{L}&=&\frac{4\pi ^{2}\alpha }{K}\left[ \frac{\left| 
\vec{q}\right| ^{2}}
{Q^{2}}W_{2}+W_{1}\right].
\end{eqnarray}

The structure functions can be calculated from
\begin{eqnarray}
W_{1} ~&& = -\frac{1}{2}\sum _{\lambda =\pm 1}(2\pi )^{6}k^{2}\int 
d\vec{p_{1}}\delta (\omega+m_{d}-E_{\pi }(\vec{k})-E_{N}
(\vec{p_{1}})-E_{N}(\vec{P_{2})}) \nonumber\\
&& \times ~\frac{1}{2J_{d}+1}\sum_{\lambda=\pm}
\sum _{M_{d},m_{s_{1}},m_{s_{2}}}
I_{m_{s_{1}},m_{s_{2}},\lambda ,M_{d}}
(\vec{k},\vec{p_{1}},\vec{p_{2}};\vec{q,}\omega),
\end{eqnarray}

\begin{eqnarray}
W_{2} ~&& = \frac{Q^{4}}{\left| \vec{q}\right| ^{4}}(2\pi )^{6}k^{2}
\int d\vec{p_{1}}\delta (\omega+m_{d}-E_{\pi }(\vec{k})-E_{N}
(\vec{p_{1}})-E_{N}(\vec{p_{2}})) \nonumber\\
&& \times ~\frac{1}{2J_{d}+1}
\sum _{M_{d},m_{s_{1}},m_{s_{2}}}
I_{m_{s_{1}},m_{s_{2}},0,M_{d}}
(\vec{k},\vec{p_{1}},\vec{p_{2}};\vec{q},\omega)
-\frac{Q^{2}}{\left| \vec{q}\right| ^{2}}W_{1},
\end{eqnarray}
where $\lambda$ is the photon polarization,
$m_{s_i}$ and $M_d$ are the z-component of the $i$-th nucleon and
the deuteron respectively. The pion momentum is
$\vec{k}$ and the momenta of the outgoing two nucleons
are $\vec{p}_1$ and $\vec{p}_2$. 
The energy variables
are defined as $E_\pi(\vec{k})=\sqrt{m_\pi^2+\vec{k}^2}$ for the pion
 and $E_N(\vec{p})=\sqrt{m_N^2+\vec{p}^2}$ for the nucleon. 
The isospin
variables are suppressed to simplify the presentation.

For the considered reaction mechanisms illustrated in Fig.\ref{fig3}, 
the total amplitude in Eqs.(7)-(8) is
\begin{eqnarray}
I_{m_{s_{1}},m_{s_{2}},\lambda ,M_{d}}
=<\vec{p}_1m_{s_{1}},\vec{p}_2m_{s_{2}};\vec{k}\mid
 I\mid q\lambda,\Psi^{J_dM_d} >, 
\end{eqnarray}
where $\Psi^{J_dM_d}$ is the deuteron wavefunction, and
\begin{eqnarray}
I=I^{(Imp)}+I^{(FSI)}+I^{(Resc)}+I^{(BB)}.
\end{eqnarray}
Explicitly, the impulse term(Fig.3a) is defined by
\begin{eqnarray}
I^{(Imp)} = \sum_{i=1,2} A(i),
\end{eqnarray}
where $A(i)$ is the pion electroproduction operator on the $i-$th nucleon.
The Nucleon-Nucleon $NN$ final state interaction term(Fig.3b) is
\begin{eqnarray}
I^{(FSI)}=\sum_{i=1,2} T_{NN,NN}(E-K_\pi) G_{NN}(E-K_\pi)A(i).
\end{eqnarray}
Here $T_{NN,NN}(\omega)$ is the $NN$ scattering operator and
$K_\pi$ is the free energy operator for the pion. The
$NN$ propagator is defined by
\begin{eqnarray}
G_{NN}(\omega)&=&\frac{1}{\omega-K_N(1)-K_N(2)+i \epsilon} ,
\end{eqnarray}
where $K_N(i)$ is the free energy operator for the $i-$th nucleon. 
The pion rescattering term(Fig.\ref{fig3}b) is defined by
\begin{eqnarray}
I^{(res)}=\sum_{i\neq j}t_{\pi N}(E-K_N(j),i)G_{\pi NN}(E)A(j),
\end{eqnarray}
where $t_{\pi N}(\omega,i)$ 
is the $\pi N$ scattering operator on the $i-$th nucleon,
 and the $\pi NN$ propagator is defined by
\begin{eqnarray}
G_{\pi NN}(E)=\frac{1}{E-K_N(1)-K_N(2)-K_\pi + i \epsilon}.
\end{eqnarray}
The baryon-baryon interaction term(Fig.\ref{fig3}d) is   
\begin{eqnarray}
I^{(BB)}=\sum_{i=1,2}\sum_{B,B'=N,\Delta}h_{\pi N, B}(i)
G_{BN}(E)T_{BN,B'N}(E)G_{B'N}(E)F_{B,\gamma N}(i),
\end{eqnarray}
where the vertex interactions $h_{\pi N, B}$ and $F_{B, \gamma N}$
describe the $B\leftrightarrow \pi N$ and $B\leftrightarrow \gamma N$
transitions respectively. In addition to $G_{NN}(E)$ defined by
Eq.(13), Eq.(16) also depends on the $\Delta N$ propagator.
 It is defined by
\begin{eqnarray}
G_{\Delta N}(E)&=&\frac{1}{E-K_N(1)-K_\Delta(2) - \Sigma_\Delta(E)},
\end{eqnarray}
where $K_\Delta$ is the free energy operator for the $\Delta$, and
 $\Sigma_\Delta(E)$ is the $\Delta$ self-energy evaluated in
the presence of a spectator nucleon. 

In our calculations, the matrix element of $A(i)$ is 
 generated from the SL model.
The matrix elements of the $\pi N$ scattering operator $t_{\pi N}$,
the baryon-baryon scattering operator  $T_{BN,B'N}$
with $B,B'=N,\Delta$ are generated from the $\pi NN$ model developed in
Ref.\cite{lm}.
Therefore, our calculations of the total amplitude Eq.(9) and
the $d(e,e'\pi^+)$ cross sections are free of adjustable parameters. 

\section{Results and Discussions}

We first consider the data from Saclay \cite{saclay} in 1990. In Fig.4, 
we show the relative importance between the four amplitudes illustrated
in Fig.\ref{fig3}. We see that the impulse term(solid curve) dominates. The
final $NN$ interaction(dashed curve) gives comparable contributions
 only at energies very close to the threshold. 
This is due to the fact that in this region the energy of the
outgoing $NN$ state is very low and the final $NN$ interaction is
dominanted by the attractive
$^1S_0$ force. The contributions from the
pion rescattering mechanism(dot-dashed curve) 
and the $BB$ mechanism(dotted curve) are clearly very weak.

In Fig.\ref{fig5}, we compare the predicted cross sections(solid curve) with
the data\cite{saclay}. We see that the general feature of the data
is reproduced. However, the discrepancy is rather significant in 
the region where the missing mass is close to 1900 MeV. This could
be due to the deficiency of the SL model or the 
higher order reaction mechanisms which are neglected
in this work. 
In the same figure, we also show the results(dashed curve)
from the impluse term only. It is seen that the $\pi NN$ interaction
mechanisms, illustrated in Figs.(\ref{fig3}b-\ref{fig3}d), 
yield an about 10 $\%$ effect in the
region near the peak position. We have found that this is mainly due to
the interference between the impulse amplitude and
the FSI amplitude. The effects due to pion rescattering and
$BB$ interactions are negligible. 

We next consider a recent experiment at JLab\cite{anl}.
The calculated cross sections from each mechanism illustrated in Fig.3
are compared in Fig.\ref{fig6}. Here 
the $BB$ interaction term(dotted curve) and the pion rescattering
term(dot-dashed curve) become comparable. However, the impulse term(solid
curve) is much larger than the other terms. Accordingly, the effects
due to the  $\pi NN$ (rescattering) mechanisms is much less. This is shown 
in Fig.\ref{fig7}.
In the same figure, we also see that the predicted cross sections are
about 20 $\%$ lower than the data.
This perhaps is
mainly due to the deficiency of the SL model in this
$Q^2=0.4$ (GeV/c)$^2$ region. However, it could be due to the
neglect of higher order reaction mechanisms.

\begin{figure}[bp]
\includegraphics{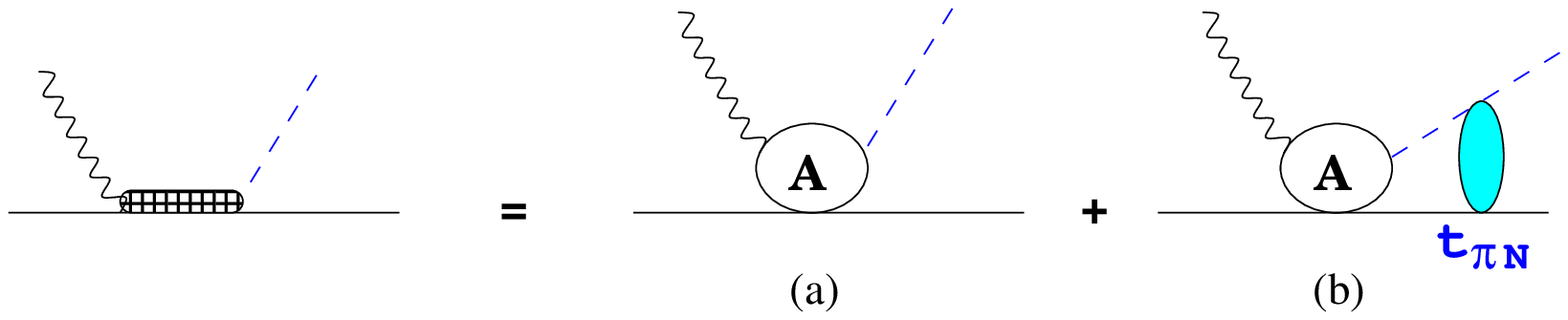}
\caption{Graphical representation of the SL elementary amplitude 
for $\pi^+$ electroproduction on the proton. Diagram (a) illustrates the
production amplitude. Diagram (b) accounts for $\pi N$ scattering.}
\label{fig1}
\end{figure}

\begin{figure}[bp]
\includegraphics{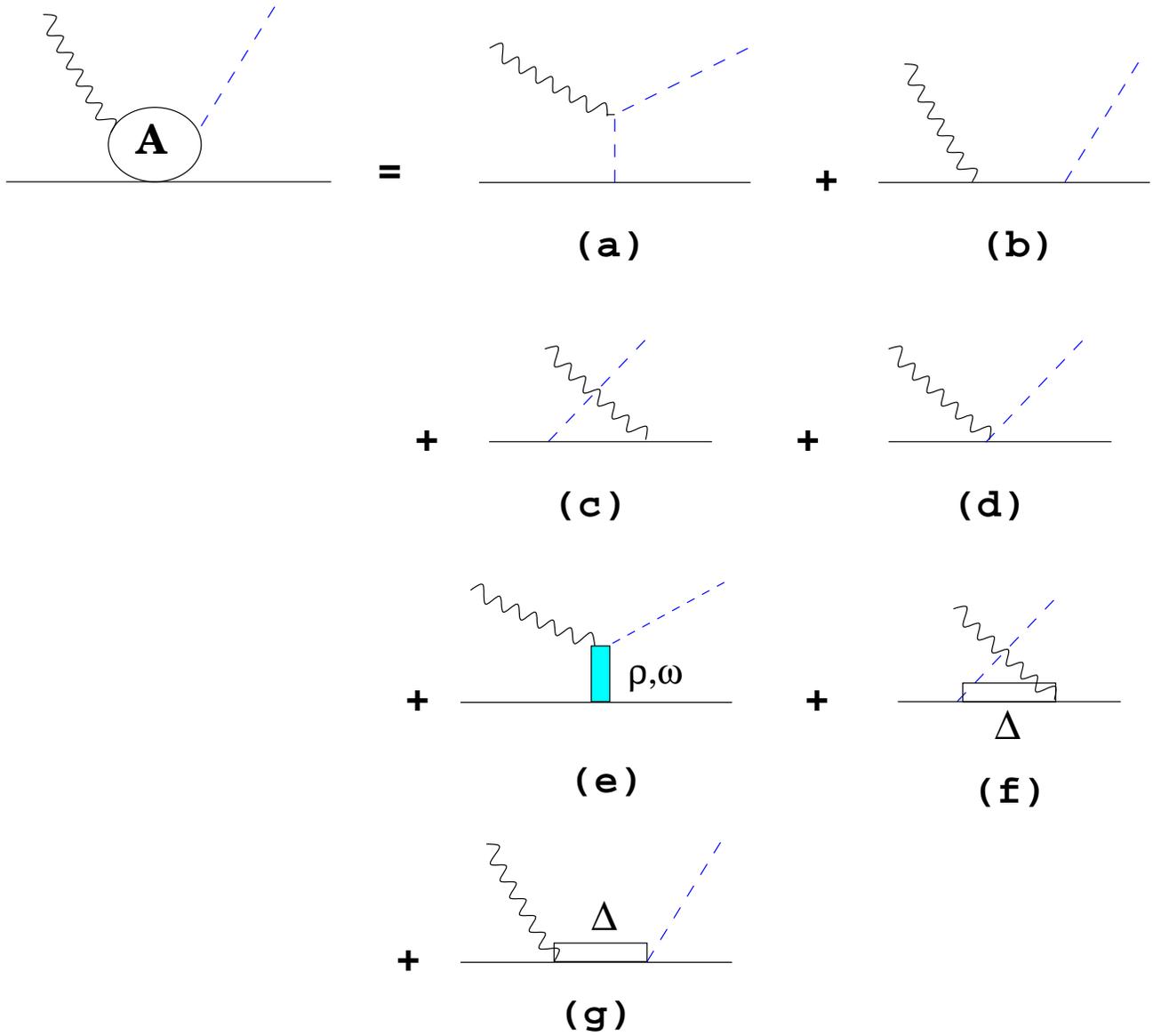}
\caption{Graphical representation of the interaction processes included in the 
SL elementary amplitude.}
\label{fig2}
\end{figure}

\begin{figure}[bp]
\includegraphics{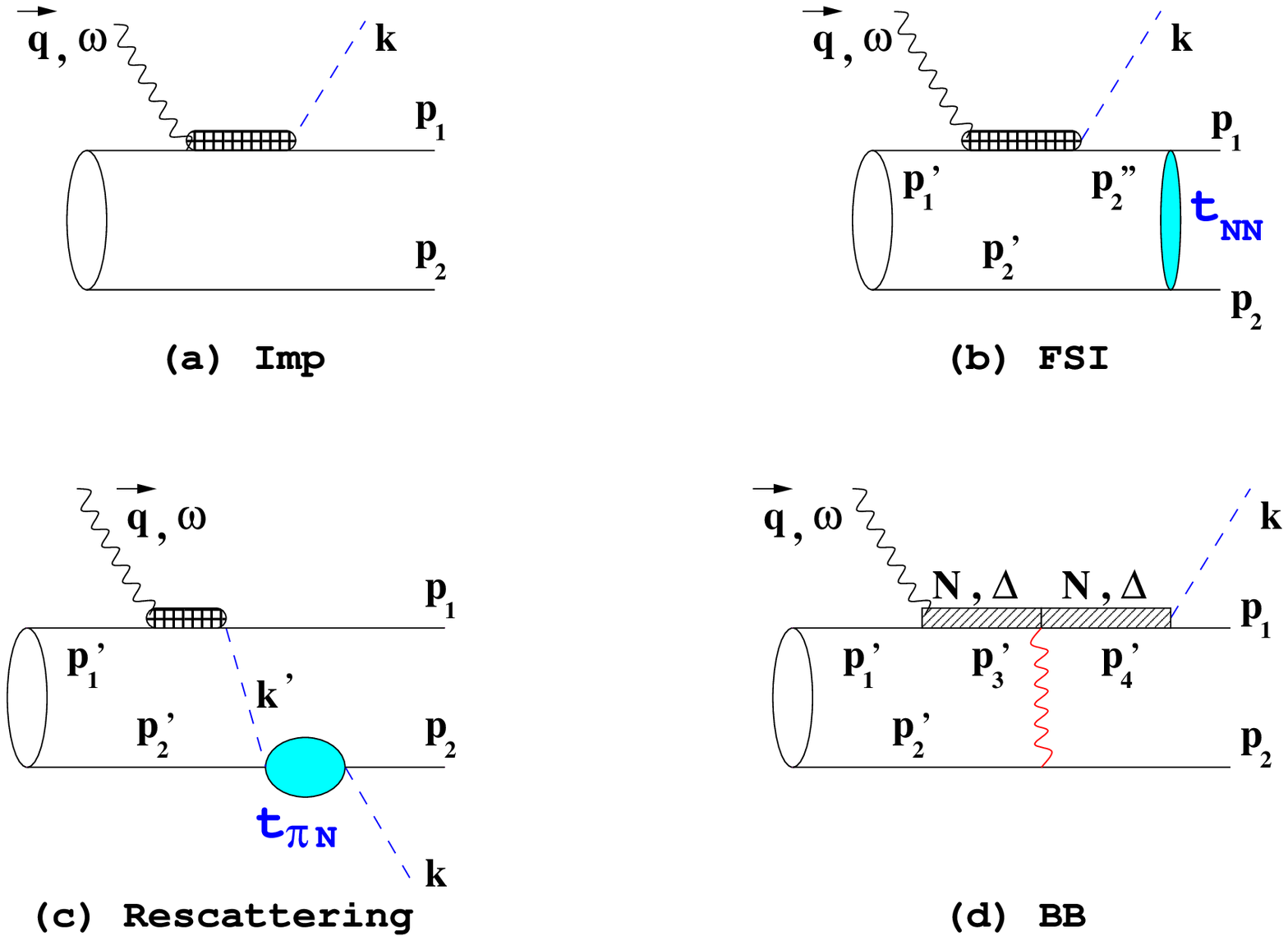}
\caption{Graphical representation of $\pi^+$ electroproduction on the deuteron Eq.(10). 
Graph (a), (b), (c) and (d) corresponds respectively to Impulse contribution Eq.(11), 
$NN$ Final State Interaction Eq.(12), pion Rescattering amplitude  Eq.(14), and 
Baryon-Baryon interaction Eq.(16).}
\label{fig3}
\end{figure}

We now turn to discussing how the measured
longitudinal cross sections can be used to learn about the
medium effects on the pion form factor and/or $\pi NN$ vertex of the
pion pole term. First we need to know the effect due to
the $\pi NN$ mechanisms.  As shown in Fig.\ref{fig8},
this effect only reduces the longitudinal cross section by about 5 $\%$.
Second, we need to know the contribution from the pion pole term. 
For the considered parallel kinematics,
the pion pole term(Fig.\ref{fig2}a) indeed dominates. On the other hand, the other
mechanisms in Figs.\ref{fig1}-\ref{fig2} 
can still have large effects through their interference
with the pion pole term. This is illustrated in Fig.\ref{fig8} where we see that
the pion pole term alone only gives about 50 $\%$ of the longitudinal cross section.
One thus must be cautious in interpreting the measured longitudinal
cross sections in terms of the medium effects on pion form factor and/or
$\pi NN$ vertex.

In summary, we have investigated the $d(e,e'\pi^+)$ reaction
for parallel kinematics by using
a dynamical model of pion electroproduction on the nucleon.
We have found that the effects
due to the final two-nucleon interactions, pion rescattering from the
second nucleon and the intermediate $NN$ and $N\Delta$ interactions
to be small.  But they
can have significant contributions
to the $d(e,e'\pi^+)$ cross sections through their interference
with the dominant impulse term. For the longitudinal cross sections,
the effects due to the interference between the
pion pole term and
other production mechanisms are also found to be
very large. Our findings indicate that these interference 
effects must be accounted for in any attempt to determine 
from the $d(e,e'\pi^+)$ data
whether the pion form factor and/or $\pi NN$ vertex of the pion pole
term are modified in the nuclear medium.
 
This work is supported by the U.S. Department of Energy, Nuclear Physics
Division, under contract No. W-31-109-ENG-38.

\newpage

\begin{figure}[bp]
\includegraphics{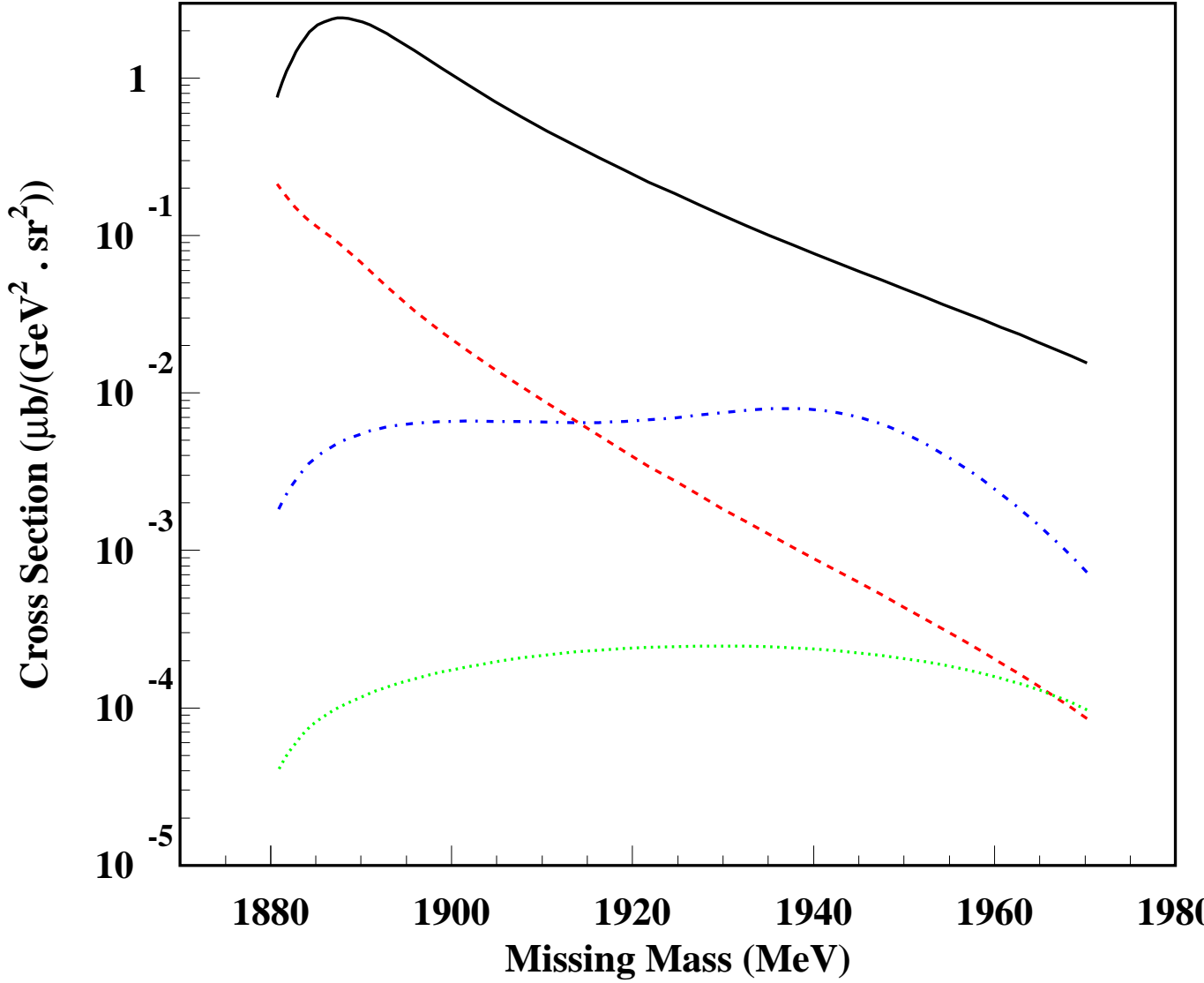}
\caption{Cross section for the processes illustrated in Fig.3 as a function 
of the missing mass $M_{x}$ defined as $M_{x} = (\omega + M_{d} - E_{\pi}(\vec{k}))^{2} 
- (\vec{q} - \vec{k})^2$. The kinematic conditions ($Q^{2} = 0.08 (GeV/c)^{2}, 
W = 1.16 GeV, E_{e} = 645 MeV, E_{e'}= 355 MeV$) are identical 
to one setting of Saclay experiment [1]. The full (dashed) curve corresponds to the 
Imp (Fig.3.a) (FSI (Fig.3.b)) contribution. The dotted-dashed (dotted) curve shows 
the rescattering (Fig.3.c) (Baryon-Baryon (Fig.3.d)) contribution.}
\label{fig4}
\end{figure}

\begin{figure}[bp]
\includegraphics{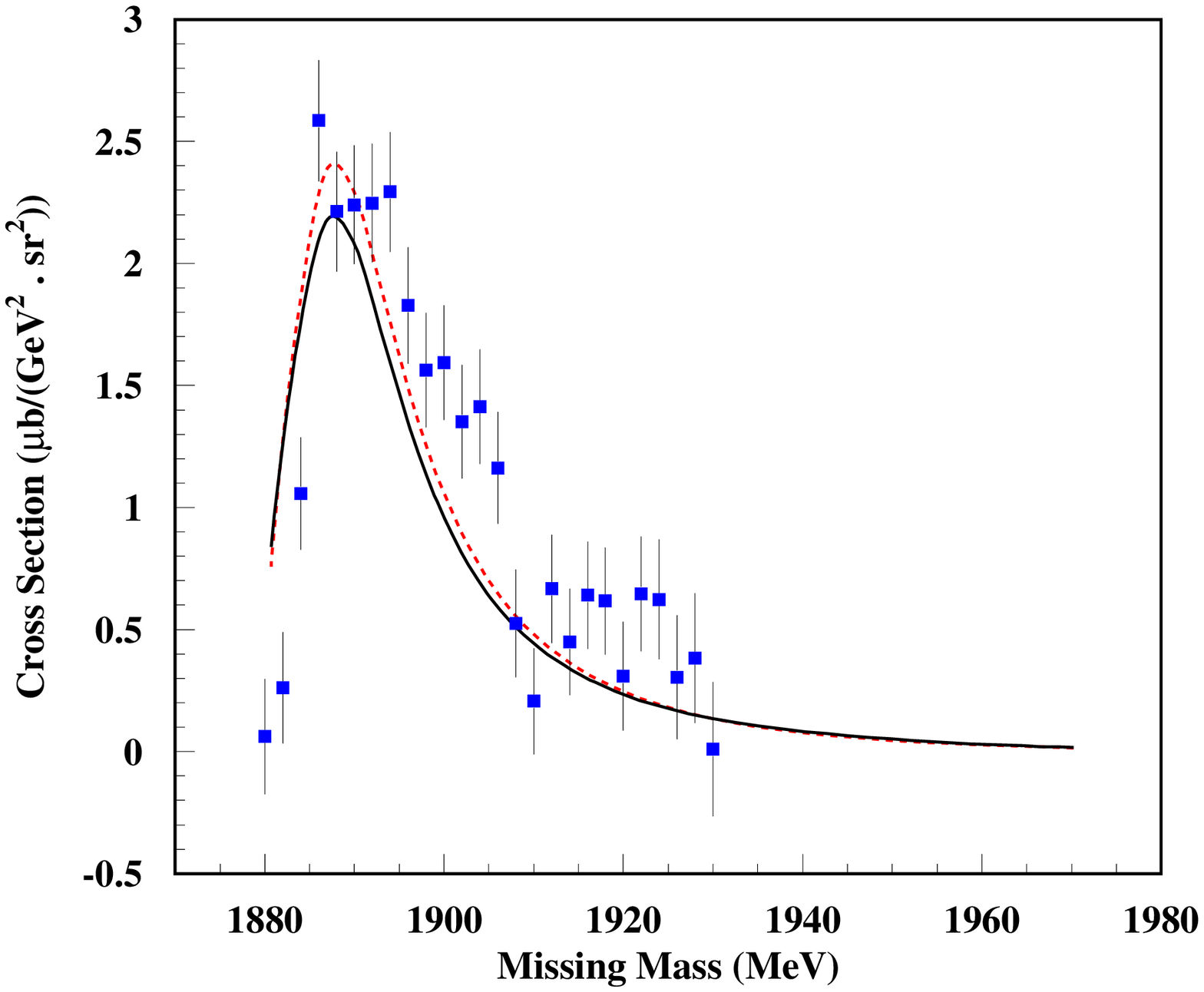}
\caption{Data on $d(e,e'\pi^+)$ cross section [1]. The kinematic conditions are 
the same as in  Fig.\ref{fig4}. The dashed curve corresponds 
to the impulse calculation (Fig.\ref{fig6}a). The solid curve shows the effect of 
inclusion of the two-body interactions (Fig.(\ref{fig3}b-\ref{fig3}c-\ref{fig3}d)).}
\label{fig5}
\end{figure}

\begin{figure}[bp]
\includegraphics{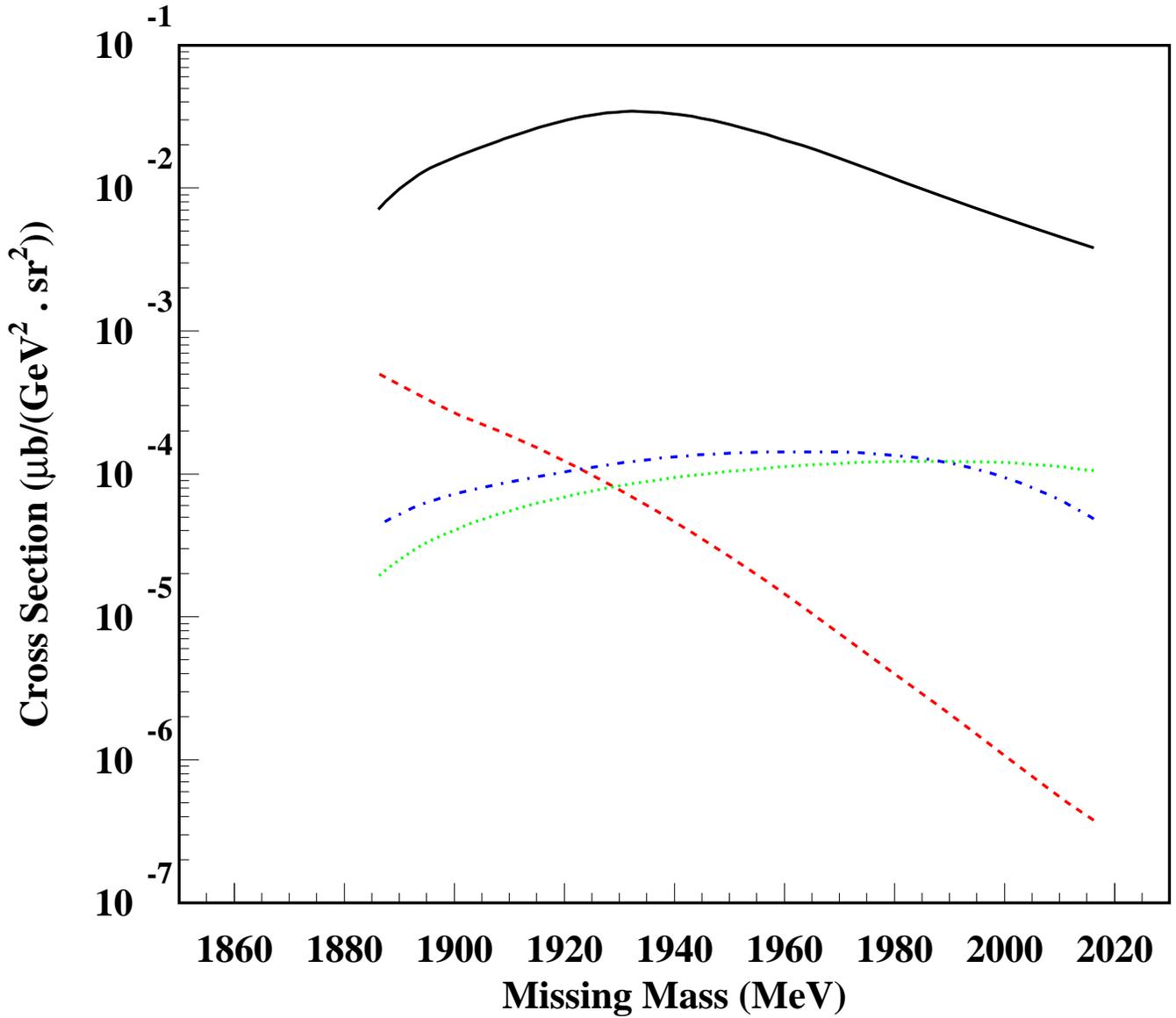}
\caption{Cross section for the processes illustrated in Fig.\ref{fig6} as a function 
of the missing mass. 
The kinematic conditions ($Q^2 = 0.4 (GeV/c)^2, W = 1.16 GeV, E_e = 844 MeV, E_{e'}= 395 MeV$) 
are identical to one setting of JLab experiment [2]. The full (dashed) curve corresponds to 
the Imp (FSI) contribution. The dotted-dashed (dotted) curve shows 
the rescattering (Baryon-Baryon) contribution.}
\label{fig6}
\end{figure}

\begin{figure}[bp]
\includegraphics{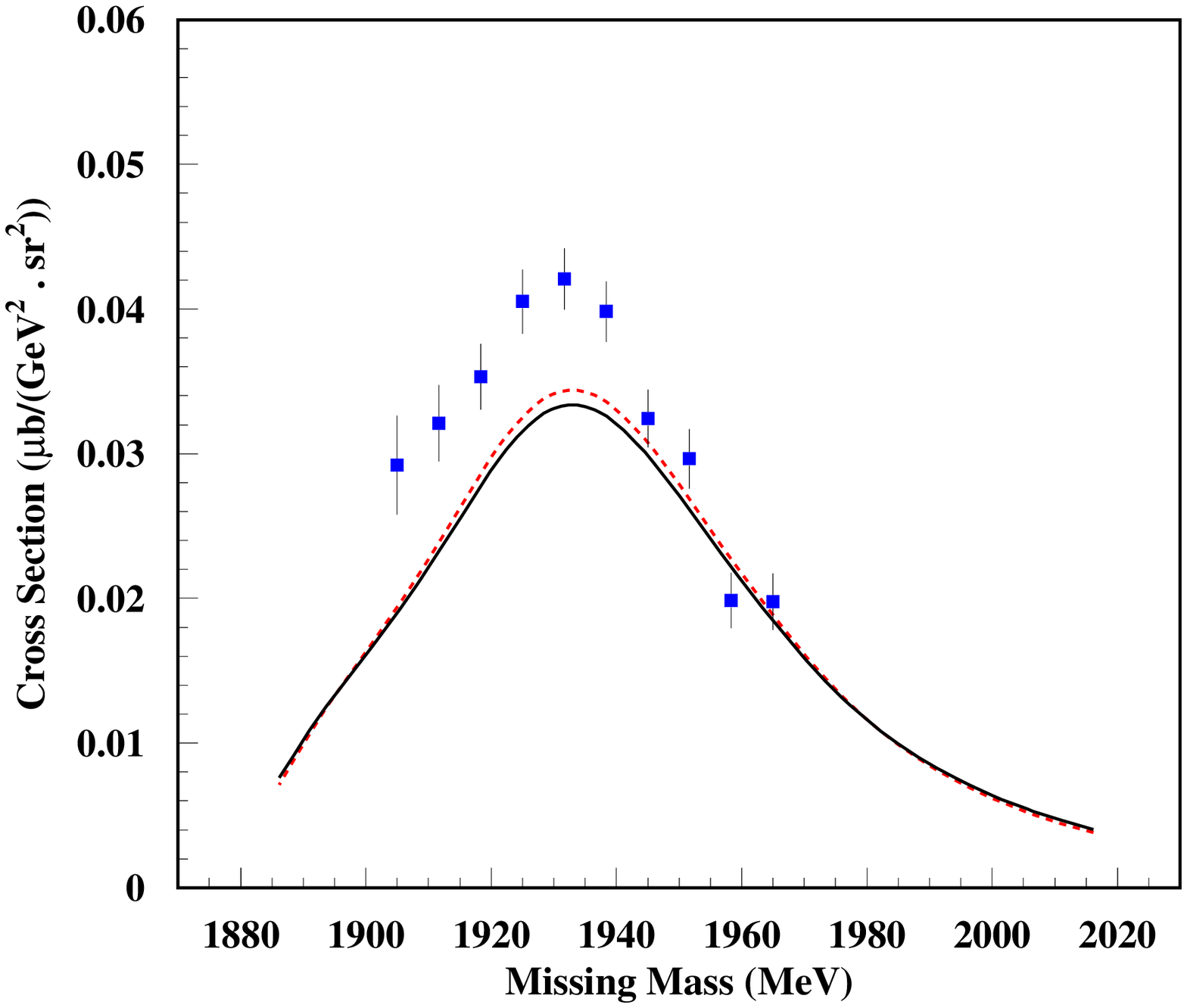}
\caption{Data on $d(e,e'\pi^+)$ cross-section [2]. The kinematic conditions are 
the same as in  Fig.\ref{fig6}. The dashed curve corresponds 
to the impulse calculation (Fig.\ref{fig6}a). The Solid curve shows the effect of inclusion 
of two-body interaction (Fig.\ref{fig3}b,c,d).}
\label{fig7}
\end{figure}

\begin{figure}[bp]
\includegraphics{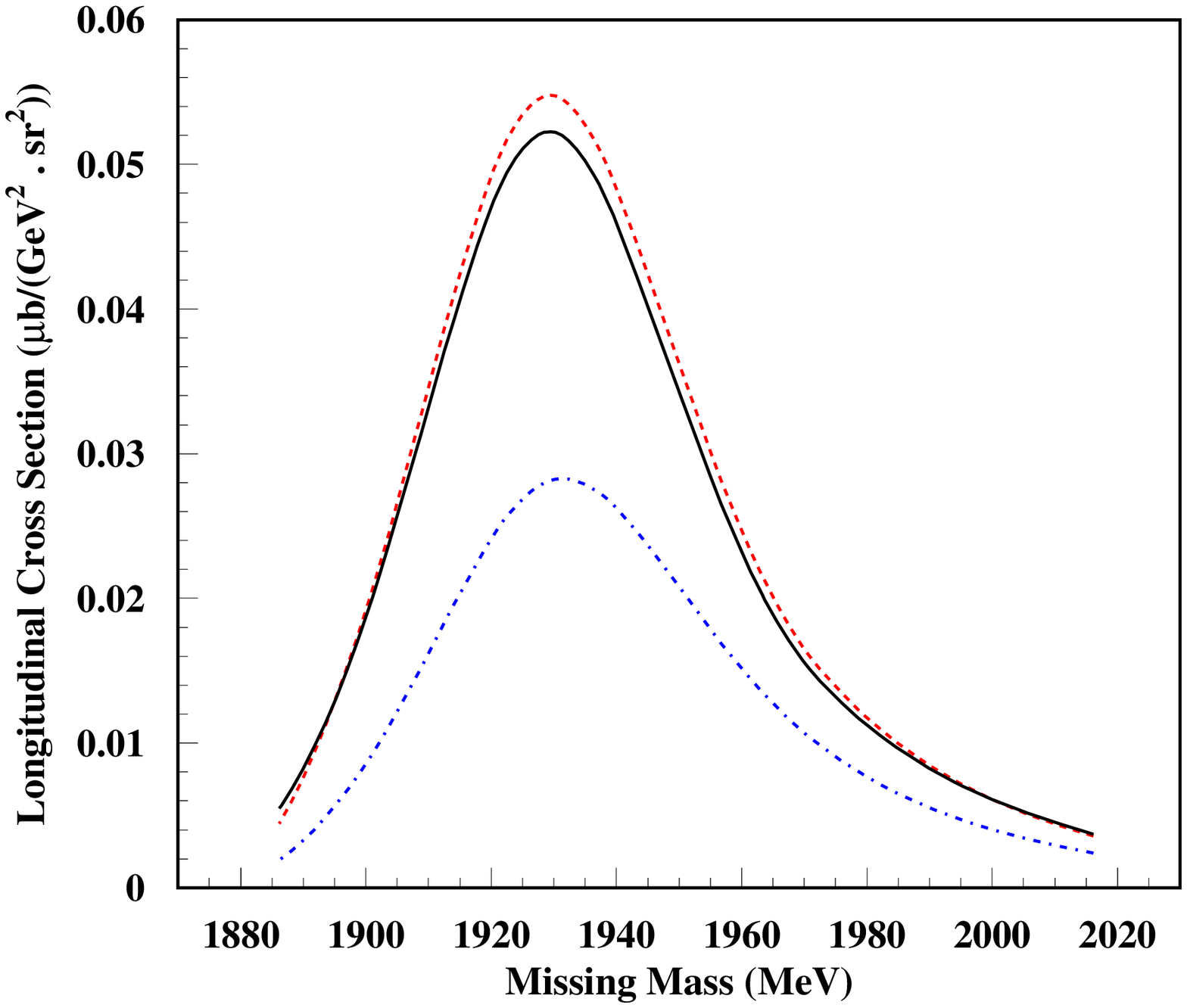}
\caption{Dashed curve corresponds to the longitudinal cross section Eq.(6) for one of the JLab 
kinematics ($Q^2 = 0.4 (GeV/c)^2, W = 1.16 GeV, E_e = 844 MeV, E_{e'}= 395 MeV$) including only 
the Impulse contribution (Fig.\ref{fig3}a). The solid curve shows the effect of inclusion of the 
two-body interaction(Fig.(\ref{fig3}b-\ref{fig3}c-\ref{fig3}d)). The dotted-dashed curve shows 
the same observable considering only the pole term (Fig.\ref{fig2}a).}
\label{fig8}
\end{figure}



\begin{references}

\bibitem{saclay}
R. Gilman et  al., Phys. Rev. Lett. {\bf 64}, 622 (1990)

\bibitem{anl}
D. Gaskell et al.,submitted to Phys. Rev. Lett (2001).

\bibitem{sl}
T.~Sato and T.-S.~H. Lee, Phys. Rev. C {\bf 54}, 2660 (1996).

\bibitem{garcilazo}
As reviewed by 
H. Garcilazo and T. Mizutani, {\it $\pi NN$ System}(World Scientific
,Singapore, 1990).

\bibitem{lm} T.-S. H. Lee and A. Matsuyama, Phys. Rev. C{\bf 36}, 1459 (1987)


\end{references}
\end{document}